\documentclass[aps,prd,a4paper,onecolumn,showpacs,showkeys,preprintnumbers,amsmath,amssymb,notitlepage]{revtex4}

\usepackage{graphicx}
\usepackage{epsfig}
\usepackage{amsmath}
\usepackage{amssymb}
\usepackage{dcolumn}

\begin{document}

\title{Minimal Scales from an Extended Hilbert Space}

\author{Martin Kober}
\email{kober@th.physik.uni-frankfurt.de}
\author{Piero Nicolini} 
\email{nicolini@th.physik.uni-frankfurt.de}
 
\affiliation{Frankfurt Institute for Advanced Studies (FIAS),
Institut f\"ur Theoretische Physik, Johann Wolfgang Goethe-Universit\"at, 
Ruth-Moufang-Strasse 1, 60438 Frankfurt am Main, Germany}
\date{\today}

\begin{abstract}
We consider an extension of the conventional quantum Heisenberg algebra, assuming that coordinates as well as momenta fulfil
nontrivial commutation relations. As a consequence, a minimal length and a minimal mass scale are implemented. Our commutators
do not depend on positions and momenta and we provide an extension of the coordinate coherent state approach to Noncommutative
Geometry. We explore, as toy model, the corresponding quantum field theory in a (2+1)-dimensional spacetime. Then we
investigate the more realistic case of a  (3+1)-dimensional spacetime, foliated into noncommutative planes. As a result, we
obtain propagators, which are finite in the ultraviolet as well as the infrared regime.
\end{abstract}

\maketitle

\section{Introduction}

In the last years there has been an explosion of interest about physics in the presence of a minimal length.
Reviews about the main approaches where a minimal length appears in the context of quantum  gravity
are given in \cite{ReviewQG}, with special reference to noncommutative geometry in \cite{ReviewNCG}.
Indeed it is expected that our universe, conventionally described in terms of a smooth differential manifold endowed
with a Riemannian structure, exhibits an uncertainty principle which prevents one from measuring positions to better accuracies
than the Planck length. The reason in support for this loss of resolution can be found in the fact that the momentum and the
energy required to perform such a measurement will itself dramatically modify the spacetime geometry at these scales
\cite{DeWitt}. In support of this general belief, there are all the current formulations of a quantum theory of gravity:
String Theory incorporates the concept of minimal length through the string tension, a fundamental parameter in the theory, while in Loop Quantum Gravity a minimal length naturally arises after quantizing the
gravitational field and formulating an area or a volume operator.
Along this line of thought, one can also implement a minimal length in spacetime by assuming nontrivial coordinate commutation
relations and thus postulating the existence of a Noncommutative Geometry as fundamental property of nature. While its original
formulation is dated back to early times \cite{Snyder:1946qz}, a model of Noncommutative Geometry has recently been obtained
within the theory of open strings, whose end points show a noncommutative behavior on D-branes \cite{Seiberg:1999vs}.
Alternatively one may assume that the conventional uncertainty principle among coordinates and momenta in Quantum Mechanics can
be generalized in order to get a modified dispersion relation at higher momenta and implement a minimal length in a quantum
theory \cite{GUP}.
More recently, the emergence of a minimal length has been considered in a variety of physical situations. The inclusion of
noncommutative effects in General Relativity has led to singularity free, thermodynamically stable black hole spacetimes
\cite{NCBHs} [for a review see \cite{review} and the references therein] and cosmological inflationary scenarios driven by
quantum geometry fluctuations rather than by an inflaton field \cite{NCcosmo}. A minimal length has been extensively
implemented to study black hole thermodynamics \cite{NCthermo}, to describe traversable wormholes sustained by noncommutative
geometry fluctuations \cite{NCwormhole} and to address the trans-planckian problem in particle production mechanisms in
accelerated reference frames \cite{NCApplications} and expanding universes \cite{NCparticle}. On the particle phenomenology
side, a lot of attention has been devoted to predict some signatures of an hypothetical production of microscopic regular black
holes \cite{NCBHs} in particle detectors due to extreme energy hadron collisions at the LHC \cite{NCpheno}. Finally, as a
warranty of theoretical consistency of effective models endowed with a minimal length, there is a recent result about the
spectral dimension of spacetime: the actual spacetime dimension $\mathbb{D}$ perceived by a random walker tends to the value
$\mathbb{D}=2$ for scales approaching the minimal length, suggesting the long sought possibility of improving the
renormalization properties of gravity at the Planck scale \cite{NCspectral}.

Against this background, there is the possibility for spacetime to be endowed not only with a minimal length but also with a
minimal momentum below which interactions, gravity included, cannot be resolved. This corresponds to some old ideas 
about a the introduction of a natural infrared (IR) cutoff \cite{Infeld}, vivified by some  recent contributions. Indeed
attempts to provide an IR modification of General Relativity have
been formulated to describe the present acceleration of the universe. As a result a new dynamics of the gravitational field has
been derived, giving a small mass to the graviton and keeping unchanged the spacetime manifold \cite{Graviton}. In the
framework of semi-classical gravity, the description of an evaporating black hole is based on high frequency field mode 
approximations to avoid unpleasant IR infinite contributions to the quantum stress energy tensor of matter fields \cite{HF}.
Instead of embarking in {\it ad hoc} approximations, it has been recently advocated the so called Ultra Strong Equivalence
Principle, according to which the vacuum expectation value of the (bare) energy momentum tensor has the same form as in flat
space, namely a quartically divergent term and no spacetime dependent (subleading) terms \cite{USEP}. If this principle is
true, we will need an IR completion of standard semi-classical gravity to get rid of all spacetime dependent subleading terms.
This fact is connected to the problem of the cosmological constant: even if only the fluctuations of the quantum vacuum energy
actually contribute to the value of the cosmological constant \cite{cosmoenergyfluc}, it has been recently underlined that
the value will still come too large without invoking an IR regulator \cite{Padmanabhan:2004qc}.

Recently, the introduction of an ultraviolet (UV) cutoff has been obtained in a clear and transparent way by averaging
noncommutative coordinate operators

\begin{eqnarray} 
\left[\hat{q}^i,\hat{q}^j\right]&=&i\Theta^{ij}
\label{NCrel}
\end{eqnarray}
on coordinate coherent states \cite{NCQFT,NCGEOM,Banerjee:2009xx}. As a result every point-like structure is smeared out by the
presence of the new quantum of length in the manifold. For instance the Euclidean propagator in coordinate space can be found
by calculating the Fourier transform of the Schwinger representation

\begin{eqnarray}
\frac{e^{-\theta \vec p^2/2}}{\vec p^2+m^2}=e^{\theta m^2/2}\int_{\theta/2}^{\infty}ds\,e^{-s(\vec p^2+m^2)},
\label{Schwinger}
\end{eqnarray}
where $\sqrt\theta$ is the minimal length implemented by (\ref{NCrel}). Coming back to the Lorentzian signature 
and considering the massless case, we find that the modified Wightman-Green function is 

\begin{equation}
G_\theta(\Delta x)=-\frac{1}{ 4\pi^2 (\Delta  t^2-\Delta \vec{x\,}^2)}\left[1-e^{-(\Delta t^2-\Delta \vec{x\,}^2)/2\theta}\right], 
\end{equation}
which is finite for any time-like interval $\Delta x$ in the coincidence limit \footnote{One must have care of the consistency
between the chosen signature and the sign of the exponent in the exponential damping factor.}.
On the other hand, the theory behaves nicely, as $G_\theta$ reduces to its conventional form in the limit $\theta\rightarrow
0$. We notice that the above smearing effect induced by the noncommutative character of the coordinates corresponds to a
modified integral measure in momentum space (\ref{Schwinger}), without invoking any deformed product among functions
\cite{ReviewNCG}. This new integral measure in momentum space has a akin in the context of the Generalized Uncertainty Principle
(GUP) approaches \cite{GUP}. Starting from an $n$ dimensional Euclidean manifold where the commutation relations are

\begin{eqnarray} 
\left[\hat{q}^i, \hat{p}_j\right]&=&i\delta^i_{j}\left(1+ f(\hat{ \vec{p} }^2)\right),\nonumber \\
\left[\hat{p}_i, \hat{p}_j\right]&=&0,
\end{eqnarray}
one obtains indeed that coordinate operators fulfil the following noncommutative relation

\begin{eqnarray} 
\left[\hat{q}^i, \hat{q}^j\right]&=&-2i\ f^\prime (\hat{ \vec{p} }^2)\left(1+f(\hat{ \vec{p} }^2)\right)\hat{L}^{ij},
\label{NCGUP}
\end{eqnarray}
where $\hat{L}_{ij}$ is the generator of rotations. The integration measure in momentum space

\begin{equation}
1=\int\frac{d^n p}{\left(1+f( \vec{p} ^2)\right)}\left|p\right>\left<p\right|
\end{equation}
becomes suppressed in the UV region, because $f(\vec{p}^2)$ depends on positive powers of the argument and usually is
$f(\vec{p}^2)=\beta {  \vec{p}}^2$. As a result the modification to the measure coming from (\ref{Schwinger}) turns out to be
even stronger, since it corresponds to the choice

\begin{equation}
f(\vec{p}^2)= e^{\theta \vec p^2/2}-1
\end{equation}
exhibiting an exponential suppression. This is in agreement with the fact that coherent states are the states of minimal
uncertainty on the noncommutative manifold, while the delocalization coming from (\ref{NCGUP}) can be arbitrarily large as the
momentum increases. For this reason, even if in the framework of GUP, both the UV and the IR cutoff has been already obtained
\cite{GUPM}, we will proceed in this paper towards an extension of the coherent state approach \cite{NCQFT} in order to endow
the spacetime of a minimal length as well as a minimal momentum/mass.

The paper is organized as follows: we start presenting the new algebra for the positions and momenta and we briefly review the
employment of coherent state for the case of a noncommutative plane (referring just to coordinates) \cite{NCQFT}. 
In Section IV, we extend the noncommutative plane case to the presence of noncommutative momenta and we obtain plane waves
endowed with damping factors, depending on both the UV and the IR cutoffs. In Section V, we derive the new form of field
functions and we calculate modified propagators, considering the four dimensional spacetime too (Section VI).
In the last section we draw the conclusions.

\section{General Assumption of noncommuting positions and momenta}

The starting point is the extension of the Heisenberg algebra for the operators describing the spacetime positions
$\hat Q^\mu$ and the corresponding momenta $\hat P_\mu$

\begin{equation}
\left[\hat Q^\mu,\hat Q^\nu\right]=i\Theta^{\mu\nu} \quad,\quad 
\left[\hat Q^\mu,\hat P_\nu\right]=i\delta^{\mu}_{\nu} \quad,\quad 
\left[\hat P_\mu,\hat P_\nu\right]=i\Sigma_{\mu\nu},
\label{Algebra}
\end{equation}
where $\Theta^{\mu\nu}$ and $\Sigma_{\mu\nu}$ are second rank antisymmetric tensors which do not depend on coordinates and
momenta. This means that the commutation relations ($\ref{Algebra}$) represent a canonical algebra. 
Besides the usual uncertainty relation between position and momenta, $\Delta Q^\mu \Delta P_\nu \geq
\frac{1}{2}\delta^\mu_\nu$, additional relations descend from ($\ref{Algebra}$), i.e.

\begin{equation}
\Delta Q^\mu \Delta Q^\nu \geq \frac{1}{2}\left|\Theta^{\mu\nu}\right|
\quad,\quad \Delta P_\mu \Delta P_\nu \geq \frac{1}{2}\left|\Sigma_{\mu\nu}\right|.
\label{relation_xx_pp}
\end{equation}
The relations ($\ref{relation_xx_pp}$) give rise to the appearance of a smallest scale for positions and momenta. Position and
momentum operators obeying the canonical algebra ($\ref{Algebra}$) can be expressed through  a couple of four-vector operators,
$\hat \alpha^\mu$ and $\hat \beta^\mu$, which fulfil the following commutation relations

\begin{equation}
\left[\hat \alpha^\mu,\hat \alpha^\nu\right]=i\Theta^{\mu\nu}\quad,\quad
\left[\hat \alpha^\mu,\hat \beta_\nu\right]=0\quad,\quad
\left[\hat \beta_\mu,\hat \beta_\nu\right]=i\Sigma_{\mu\nu},
\label{algebra_alpha_beta}
\end{equation}
if there are further operators $\hat q^\mu$ and $\hat p_\mu$ which behave like usual position and momentum operators

\begin{equation}
\left[\hat q^\mu,\hat q^\nu\right]=0\quad,\quad 
\left[\hat q^\mu,\hat p_\nu\right]=i\delta^{\mu}_{\nu} \quad,\quad 
\left[\hat p_\mu,\hat p_\nu\right]=0.
\label{commutation_qk}
\end{equation}
According to ($\ref{algebra_alpha_beta}$) and ($\ref{commutation_qk}$) the position and momentum operators $\hat Q^\mu$ and
$\hat P_\mu$ can be expressed as follows

\begin{equation}
\hat Q^\mu=\hat q^\mu+\hat \alpha^\mu \quad,\quad \hat P_\mu=\hat p_\mu+\hat \beta_\mu.
\label{representation_operators}
\end{equation}
This requires that the following relations have to be satisfied

\begin{eqnarray}
\left[\hat \alpha^\mu,\hat q^\nu\right]+\left[\hat q^\mu,\hat \alpha^\nu\right]=0\quad,\quad
\left[\hat \beta_\mu,\hat p_\nu\right]+\left[\hat p_\mu,\hat \beta_\nu\right]=0\quad,\quad
\left[\hat \alpha^\mu,\hat p_\nu\right]+\left[\hat q^\mu,\hat \beta_\nu\right]=0.
\label{alphabeta}
\end{eqnarray}

\section{The noncommutative plane and modifications of plane waves}

We start recalling the introduction of a UV cutoff due to the noncommutative character of coordinate operators along the lines
of \cite{NCQFT}. The core of the treatment lies in the possibility of defining a set of quasi classical coordinates, averaging
the corresponding quantum operators on suitable coherent states. The commutation relations concern the case of a
($2+1$)-dimensional manifold, with spatial noncommutative coordinates and commutative time, namely

\begin{equation}
\left[\hat Q^i,\hat Q^j\right]=i\Theta^{ij} \quad,\quad 
\left[\hat Q^i,\hat P_j\right]=i\delta^{i}_{j} \quad,\quad 
\left[\hat P_i,\hat P_j\right]=0,
\label{algebra_commuting_momenta}
\end{equation}
where

\begin{equation}
\Theta=\left(\begin{matrix}0&0&0\\0&0&\theta\\0&-\theta&0\end{matrix}\right)
\end{equation}
is a constant antisymmetric matrix.
The above relations imply that the operator $\hat Q^0$ commutes with the other coordinate operators. 
One starts by defining the following ladder operators

\begin{eqnarray}
\hat A=\frac{1}{\sqrt{2\theta}}(\hat Q^1+i\hat Q^2) \quad,\quad \hat A^{\dagger}=\frac{1}{\sqrt{2\theta}}(\hat Q^1-i\hat Q^2)
\end{eqnarray}
satisfying the commutation relation

\begin{eqnarray}
\left[\hat A,\hat A^{\dagger}\right]=1, 
\end{eqnarray}
which corresponds to the algebra of usual bosonic creation and annihilation operators. As a result the noncommutative
configuration space is isomorphic to the the Fock space spanned by the states 
$|n\rangle_\theta\equiv \frac{1}{\sqrt{n!}}\left(A^\dagger\right)^n|0\rangle_\theta$

\begin{equation}
{\cal H}_\theta=\mathrm{Span}\{|n\rangle_\theta\}_{n=0}^{\infty},
\end{equation}
where the ground state $|0\rangle_\theta$ is the state annihilated by $A$.
To distinguish the conventional quantum phase space from the above Fock space, we will introduce the notation

\begin{equation}
{\cal H}_\hbar=\mathrm{Span}\{|n\rangle_\hbar\}_{n=0}^{\infty},
\end{equation}
where the states $|n\rangle_\hbar$ are built starting from the commutation relation $\left[\hat q^\mu,\hat
p_\nu\right]=i\delta^{\mu}_{\nu}$ between coordinates and momenta. As stressed in \cite{Banerjee:2009xx}
Quantum Mechanics works on ${\cal H}_\hbar$, while Noncommutative Geometry works on ${\cal H}_\theta$. In both cases
observables are defined as self-adjoint operators acting respectively on ${\cal H}_\hbar$ and ${\cal H}_\theta$ and they form
algebras denoted as ${\cal A}({\cal H}_\hbar)$ and ${\cal A}({\cal H}_\theta)$. In analogy to Quantum Mechanics one can
introduce an ``Hamiltonian'' $\hat H_\theta\in {\cal A}({\cal H}_\theta)$ as 

\begin{equation}
\hat H_\theta\equiv \frac{(\hat Q^1)^2}{2}+\frac{ (\hat  Q^2)^2}{2}=\theta\left(\hat N_\theta +\frac{1}{2}\right),
\end{equation}
where the corresponding ``number´´ operator $\hat N_\theta=A^\dagger A$ identifies the raising and lowering operators

\begin{equation}
\left[\hat N_\theta,\hat A\right]=-\hat A \quad,\quad \left[\hat N_\theta,\hat A^\dagger \right]=+\hat A^\dagger . 
\end{equation}
It is straightforward to define coherent states $|a\rangle_\theta$ as eigenstates of $A$  according to 

\begin{equation}
\hat A|a\rangle_\theta=a|a\rangle_\theta,
\end{equation}
where

\begin{equation}
|a\rangle_\theta =e^{-{\frac{|a|^2}{2}}}\sum_{n=0}^{\infty}{\frac{a^n}{\sqrt{n!}}}|n\rangle_\theta
=e^{-{\frac{|a|^2}{2}}}e^{a\hat A^\dagger}|0\rangle_\theta .
\end{equation}
To calculate the modification to the plane waves due to the presence of noncommutative coordinate operators we need to recall
that $e^{ip_1q^1+ip_2q^2}$ is a representation of the eigenstates of the momentum operator $\hat p_i$, namely

\begin{equation}
\hat p_i\left|p_i\right>=p_i\left|p_i\right>,
\end{equation}
with 
\begin{equation}
\psi_\hbar\equiv\left<q^i|p_i\right>=e^{ip_iq^i}.
\label{usual_plane_waves}
\end{equation}
This is true only if we work in the framework of conventional Quantum Mechanics. Indeed $p_i$ and $q^i$ appear as usual numbers
within the expressions for the plane waves and represent the corresponding eigenvalues of the states.
In the Noncommutative Geometry case  the quantities appearing in the expression of the plane waves which correspond to
$\hat Q^i$, are still operators, since we have an additional Fock space ${\cal H}_\theta$. They behave like usual numbers with
respect to the usual Hilbert space of Quantum Mechanics ${\cal H}_\hbar$, but they are operators with respect to the additional
Hilbert space ${\cal H}_\theta$. Therefore it is necessary to define a set of classical coordinates, which are usual numbers
with respect to the complete Hilbert space $\mathcal{H}=\mathcal{H}_\hbar \otimes \mathcal{H}_\theta$.
This can be done by averaging coordinate operators on the above coherent states, which turn out to be the states of minimal
uncertainty on the noncommutative plane

\begin{equation}
Q^1\equiv {}_\theta\langle a|\hat Q^1 |a\rangle_\theta = \sqrt{2\theta}\ \mathrm{Re}\ a\quad,\quad 
Q^2\equiv {}_\theta\langle a|\hat Q^2 |a\rangle_\theta = \sqrt{2\theta}\ \mathrm{Im}\ a.
\label{coordinates}
\end{equation} 
According to ($\ref{representation_operators}$) we can represent the coordinate operators as

\begin{equation}
\hat Q^i=\hat q^i+\hat \alpha^i.
\end{equation}
It is important to emphasize that conventional coordinate operators $\hat q^i$
have no influence on the Fock space ${\cal H}_\theta $ because they fulfil the relation 
$[\hat \alpha^i,\hat q^j]+[\hat q^i,\hat \alpha^j]=0$ of equation ($\ref{alphabeta}$).
Since in the usual expressions for plane waves ($\ref{usual_plane_waves}$) there are eigenvalues $q^i$ and $p_i$ of
the usual operators $\hat q^i$ and $\hat p_i$ and not the operators themselves, in the generalized case of this paper it is
necessary to define quantities $\mathcal{Q}^i$ which are eigenvalues with respect to the Hilbert space of usual quantum
mechanics $\mathcal{H}_{\hbar}$ but operators with respect to the additional Hilbert space $\mathcal{H}_{\theta}$ 

\begin{eqnarray}
\mathcal{Q}^i &=& q^i \hat 1+\hat \alpha^i,
\end{eqnarray}
where $q^i$ is no longer an operator. Therefore the Fock space ${\cal H}_\theta $ refers to the noncommutative character
of $\hat \alpha^\mu$ only. Now we have that the coordinates (\ref{coordinates})

\begin{equation}
Q^1=q^1+{}_\theta\langle a|\hat \alpha^1 |a\rangle_\theta = \sqrt{2\theta}\ \mathrm{Re}\ a
\quad,\quad Q^2=q^2+{}_\theta\langle a|\hat \alpha^2 |a\rangle_\theta = \sqrt{2\theta}\ \mathrm{Im}\ a
\end{equation}
turn out to be the sum of the conventional classical coordinates $q^i$ and the mean values of the noncommutative fluctuations
$\langle a|\hat \alpha^i|a \rangle$. Since we are not yet considering the complete algebra ($\ref{Algebra}$) but the simpler
case of ($\ref{algebra_commuting_momenta}$), the  momenta $\hat P_i$ are still assumed to commute with each other and of course
they act as usual numbers on ${\cal H}_\theta $. This implies that the corresponding quantities appearing in the expression for
the extended plane waves, using an analogue representation to $\mathcal{Q}^i$ have the form $\mathcal{P}_i=P_i\hat 1$. Therefore for the operator

\begin{eqnarray}
\hat A =\frac{1}{\sqrt{2\theta}}[( q^1+i q^2)\hat1+\hat \alpha^1+i\hat \alpha^2],
\end{eqnarray} 
we have the relations $\left[\mathcal{P}_i,\hat A \right]=0$ meaning that 

\begin{equation}
P_1={}_\theta\langle a|\mathcal{P}_1 |a\rangle_\theta, \quad\quad P_2={}_\theta\langle a|\mathcal{P}_2 |a\rangle_\theta.
\end{equation}
Thus if we consider as new expression for plane waves

\begin{eqnarray}
{}_\theta\langle a|e^{i\mathcal{P}_1 \mathcal{Q}^1+i\mathcal{P}_2\mathcal{Q}^2} |a\rangle_\theta,
\end{eqnarray}
where now the symbol ``$\ \hat{}\ $'' indicates an operator acting only on ${\cal H}_\theta$, we will have 

\begin{eqnarray}
{}_\theta\langle a|e^{i\mathcal{P}_1\mathcal{Q}^1+i\mathcal{P}_2\mathcal{Q}^2}|a\rangle_\theta
={}_\theta\langle a|e^{i P_1 \mathcal{Q}^1+i P_2 \mathcal{Q}^2} |a\rangle_\theta
={}_\theta\langle a|e^{i\sqrt{\theta}\hat A^\dagger P_+ +i\sqrt{\theta}\hat AP_-} |a\rangle_\theta
\label{NCplanewave}
\end{eqnarray}
with $P_\pm=\frac{1}{\sqrt{2}}(P_1\pm iP_2)$. By means of the Baker Campbell Hausdorff formula one obtains

\begin{equation}
{}_\theta\langle a|e^{i\sqrt{\theta}\hat A^\dagger P_+ +i\sqrt{\theta}\hat AP_-} |a\rangle_\theta
={}_\theta\langle a|e^{i\sqrt{\theta}\hat A^\dagger P_+}
e^{i\sqrt{\theta}\hat AP_-}e^{-(1/2)\theta P_+P_-[\hat A, \hat A^\dagger]} |a\rangle_\theta
=e^{iP_1 Q^1+iP_2 Q^2}e^{-\frac{\theta}{4}(P^2_1+P^2_2)}.
\end{equation}
As a result, we have that according to \cite{NCQFT} the plane wave is damped by an exponential factor coming from the
noncommutativity of coordinate operators,  which means that the plane wave expression is modified according to

\begin{eqnarray}
\psi_{\hbar}=e^{i \vec p\cdot \vec q}\longrightarrow \psi_{\hbar\theta}=e^{i \vec p\cdot \vec q}e^{-\frac{\theta}{4} \vec p^2}.
\end{eqnarray}
This corresponds to having a response from the noncommutative manifold which provides a natural UV cutoff for higher momenta
modes as a consequence of the loss of space resolution at short distances.

\section{Incorporation of noncommutative momenta}

We proceed now with the extension of the above procedure for a ($2+1$)-dimensional manifold to the case in which momentum
operators do not commute anymore as well as the position operators, i.e.

\begin{equation}
\left[\hat Q^i,\hat Q^j\right]=i\Theta^{ij} \quad,\quad 
\left[\hat Q^i,\hat P_j\right]=i\delta^{i}_{j} \quad,\quad 
\left[\hat P_i,\hat P_j\right]=i\Sigma_{ij},
\end{equation}
where $\Theta^{ij}$ and $\Sigma_{ij}$ are antisymmetric matrices which do not depend on the coordinates or the momenta.
For later convenience we assume the following form of nonvanishing commutators 

\begin{equation}
[\hat Q^1,\hat Q^2]=i \theta \quad,\quad [\hat P_2, \hat P_1]=i \sigma,
\label{algebra_two_dimensions}
\end{equation}
where we have set 

\begin{equation}
\Theta=\left(\begin{matrix}0&0&0\\0&0&\theta\\0&-\theta&0\end{matrix}\right)\quad,\quad
\Sigma=\left(\begin{matrix}0&0&0\\0&0&-\sigma\\0&\sigma&0\end{matrix}\right).
\label{algebra_two_dimensions_matrices}
\end{equation}
Along the line of reasoning followed for the noncommutative plane, we introduce

\begin{equation}
\hat A= \frac{1}{\sqrt{2\theta}}(\hat Q^1+i\hat Q^2) \quad,\quad \hat A^{\dagger}
=\frac{1}{\sqrt{2\theta}}(\hat Q^1-i\hat Q^2), \quad,\quad
\hat B=\frac{1}{\sqrt{2\sigma}}(\hat P_2+i\hat P_1) \quad,\quad \hat B^{\dagger}
=\frac{1}{\sqrt{2\sigma}}(\hat P_2-i\hat P_1),
\end{equation}
satisfying the commutation relations

\begin{eqnarray}
\left[\hat A,\hat A^{\dagger}\right]=1 \quad,\quad \left[\hat B,\hat B^{\dagger}\right]=1.\nonumber\\
\label{commutation_AB}
\end{eqnarray}
Again both, $\hat A$ and $\hat B$, define ``Hamiltonian'' operators as

\begin{equation}
\hat H_\theta\equiv \frac{ (\hat Q^1)^2}{2}+\frac{ (\hat Q^2)^2}{2}=\theta\left(\hat N_q +\frac{1}{2}\right)\quad,\quad
\hat H_\sigma\equiv \frac{\hat P_1^2}{2}+\frac{\hat P_2^2}{2}=\sigma\left(\hat N_p +\frac{1}{2}\right),
\end{equation}
where the corresponding ``number'' operators $\hat N_\theta=A^\dagger A$ and $\hat N_\sigma=B^\dagger B$ identify raising and
lowering operators

\begin{equation}
\left[\hat N_\theta,\hat A \right]=-\hat A \quad,\quad \left[\hat N_\theta,\hat A^\dagger \right]
=\hat A^\dagger\quad,\quad\left[\hat N_\sigma,\hat B \right]=
-\hat B \quad\quad \left[\hat N_\sigma,\hat B^\dagger\right]=\hat B^\dagger.
\end{equation}
Since now we are endowed not only with a quantum of action $\hbar$ and a quantum of length $\sqrt{\theta}$ but also with a
quantum of momentum $\sqrt{\sigma}$, we have three corresponding Fock spaces ${\cal H}_\hbar $, ${\cal H}_\theta $ and

\begin{equation}
{\cal H}_\sigma=\mathrm{Span}\{|n\rangle_\sigma\}_{n=0}^{\infty},
\end{equation}
where $|n\rangle_\sigma\equiv \frac{1}{\sqrt{n!}}B^\dagger|0\rangle_\sigma$.
 We can follow the same procedure as above and consider the following representation of the operators
$\mathcal{Q}^i\in{\cal H}_\theta$

\begin{equation}
\mathcal{Q}^i=q^i\hat 1+\hat\alpha^i
\label{coordinates_plane_waves}
\end{equation}
and $\mathcal{P}_i\in{\cal H}_\sigma $

\begin{equation}
\mathcal{P}_i=p_i\hat 1+\hat\beta_i.
\label{momenta_plane_waves}
\end{equation}
provided that $[\hat \alpha^i,\hat q^j]+[\hat q^i,\hat \alpha^j]=[\hat \beta_i,\hat p_j]+[\hat p_i,\hat \beta_j]=0$ as in ($\ref{alphabeta}$). This means that the $\mathcal{Q}^i$ are operators with respect to the additional Hilbert
space ${\cal H}_\theta$ and the $\mathcal{P}_i$ are operators with respect to the additional Hilbert space ${\cal H}_\sigma$.
However both, the $\mathcal{Q}^i$ and the $\mathcal{P}_i$, as they appear in the plane wave expression, act as usual numbers on
${\cal H}_\hbar$. Therefore the coordinates (\ref{coordinates}) read

\begin{equation}
Q^1=q^1+{}_\theta\langle a|\hat \alpha^1 |a\rangle_\theta = \sqrt{2\theta}\ \mathrm{Re}\ a\quad,\quad
Q^2=q^2+{}_\theta\langle a|\hat \alpha^2 |a\rangle_\theta = \sqrt{2\theta}\ \mathrm{Im}\ a\\
\end{equation}
and the momenta read

\begin{equation}
P_1=p_1+{}_\sigma\langle b|\hat \beta_1 |b\rangle_\sigma = \sqrt{2\sigma}\ \mathrm{Im} \ b,\quad,\quad
P_2=p_2+{}_\sigma\langle b|\hat \beta_2 |b\rangle_\sigma = \sqrt{2\sigma}\ \mathrm{Re}\ b,\\
\end{equation}
where the coherent states $|a\rangle_\theta$ and $|b\rangle_\sigma$ are eigenstates of the operators 

\begin{equation}
\hat A=\frac{1}{\sqrt{2\theta}}[(q^1+iq^2)\hat 1+\hat \alpha^1+i\hat \alpha^2]\quad,\quad 
\hat B=\frac{1}{\sqrt{2\sigma}}[(p^1+ip^2)\hat 1+\hat \beta^1+i\hat \beta^2].
\end{equation}
Since $\left[\hat \alpha^\mu,\hat \beta_\nu\right]=0$, we can conclude that ${\cal H}_\theta\cong {\cal H}_\sigma$ are
isomorphic, being $\left[\hat A,\hat B\right]=0$ and admit common eigenstates $\left. |a\ b\right>_{\theta\sigma}$.
We are now ready to compute the modifications to plane waves due to the presence of both, noncommutative positions and momenta,
corresponding to the parameters $\sqrt{\theta}$ and $\sqrt{\sigma}$.

\begin{eqnarray}
{}_{\theta\sigma}\langle a\ b |\exp\left[i\mathcal{P}_1 \mathcal{Q}^1+i\mathcal{P}_2 \mathcal{Q}^2\right]
|a \ b\rangle_{\theta\sigma}
&=&{}_{\theta\sigma}\langle a\ b |\exp\left[-t \hat A^{\dagger}\hat B^\dagger + t\hat A \hat B\right]
|a \ b\rangle_{\theta\sigma},
\label{plane_waves_AB}
\end{eqnarray}
where the dimensionless parameter $t=\sqrt{\theta\sigma}$ is the product of the quantum of length and the quantum of momentum.
Therefore we can safely assume that $t\ll 1$. Again we stress that now the symbol ``$\ \hat{}\ $'' indicates an operator only
with respect to ${\cal H}_\sigma$ and ${\cal H}_\theta$, spanned by $\{|a \ b\rangle_{\theta\sigma}\}$. To evaluate the
expression ($\ref{plane_waves_AB}$), we can use the Baker Campbell Hausdorff formula 

\begin{equation}
e^{tX} e^{tY}=e^{tX+tY+\frac{t^2}{2}[X,Y]+\frac{t^3}{12}[X,[X,Y]]-\frac{t^3}{12}[Y,[X,Y]]+...}
\label{BCH}
\end{equation}
with $X=-\hat A^{\dagger}\hat B^{\dagger}$ and $Y=\hat A \hat B$. The above formula can be written in terms of the following
combinatoric expansion 

\begin{equation}
e^{t(X+Y)}= e^{tX}~ e^{tY} ~e^{-\frac{t^2}{2} [X,Y]} ~ e^{\frac{t^3}{6}(2[Y,[X,Y]]+ [X,[X,Y]] )}
~ e^{\frac{-t^4}{24}([[[X,Y],X],X] + 3[[[X,Y],X],Y] + 3[[[X,Y],Y],Y]) } \cdots
\end{equation}
provided $t\ll 1$. We notice that 

\begin{eqnarray}
&&[X,Y]=\hat A^{\dagger}\hat A +\hat B^{\dagger}\hat B, \quad\quad [X,[X,Y]]=-2X,  \quad\quad [Y,[X,Y]]=2Y 
\end{eqnarray} 
and thus we can see that any odd term will be a linear combination of $t^{2n+1} X$ and $t^{2n+1} Y$ and every even term will be
$\sim t^{2n}[X,Y]$. Applying recursively the Baker Campbell Hausdorff formula to each term, we get

\begin{equation}
e^{t(X+Y)}= e^{(t+{\cal O}(t^3))X}~ e^{(t+{\cal O}(t^3))Y} ~e^{-(\frac{t^2}{2}+{\cal O}(t^4)) [X,Y]}.
\end{equation}
As a result, we can safely keep just the leading order in the expansion 

\begin{equation}
e^{t(X+Y)}\simeq e^{tX}e^{tY}e^{-\frac{t^2}{2}[X,Y]}
\label{approximation_BCH}
\end{equation}
and write the plane wave as 

\begin{eqnarray}
&&{}_{\theta\sigma}\langle a\ b |\exp\left[i\mathcal{P}_1 \mathcal{Q}^1+i\mathcal{P}_2 \mathcal{Q}^2\right]
|a \ b\rangle_{\theta\sigma}
={}_{\theta\sigma}\langle a\ b |\exp\left[-t \hat A^{\dagger}\hat B^\dagger
+t\hat A \hat B\right]|a \ b\rangle_{\theta\sigma}=\nonumber\\ 
&&={}_{\theta\sigma}\langle a\ b |\exp\left[-t \hat A^{\dagger}\hat B^\dagger\right]\exp\left[ t\hat A \hat B\right]
\exp\left[-(t^2/2)(\hat A^{\dagger}\hat A +\hat B^{\dagger}\hat B)\right]|a \ b\rangle_{\theta\sigma}=\nonumber
\\ &&=\exp\left[-t a^{\ast} b^\ast\right]\exp\left[ t a b\right]
\exp\left[-t^2( |a|^2 + |b|^2)/2\right]=\nonumber
\\ &&=\exp\left[i\mathcal{P}_1\mathcal{Q}^1+i\mathcal{P}_2\mathcal{Q}^2\right]
\exp\left\{-t^2\left[ \sigma (Q^1)^2+\sigma (Q^2)^2+\theta P_1^2+\theta P_2^2\right]/4\right\}
\end{eqnarray}
In summary the modifications to the conventional plane waves are of the form

\begin{equation}
\psi_\hbar = e^{i \vec p\cdot \vec q}\longrightarrow \psi_{\hbar\theta\sigma}
=e^{i \vec p\cdot \vec q-(\sigma/4) \vec q^2 -(\theta/4) \vec p^2}
\label{ampl}
\end{equation}
exhibiting both  UV and IR damping terms. This result is in agreement with what has been found within the framework of a the
Generalized Uncertainty Principle, once a relation like

\begin{eqnarray} 
\left[\hat{q}^i, \hat{p}_j\right]&=&i\delta^i_{j}\left(1+ f(\hat{ \vec{p} }^2)+ g(\hat{ \vec{q} }^2)\right)
\end{eqnarray}
is assumed as in \cite{GUPM}. As a comment we have to remember that the above result has been obtained up to subleading terms
$\sim e^{i\sigma\theta\vec q\cdot\vec p}$, which are responsible of a modified dispersion relation and are extremely suppressed
in the regime $\theta\sigma\ll 1$.

\section{Derivation of the propagator of a quantum field in a (2+1)-dimensional spacetime}

We are now ready to define a quantum field.  
 As shown in \cite{NCQFT},  the modifications to plane waves due to the employment of coherent state mean coordinates do not affect the conventional
algebra of creation and annihilation operators of quantum fields, since the modification of commutation relations
($\ref{Algebra}$) refers to the Hilbert space of single particles which is extended according to the above
transition $\mathcal{H}_\hbar \rightarrow \mathcal{H}_\hbar \otimes \mathcal{H}_\theta \otimes \mathcal{H}_\sigma$. Therefore the construction of a Fock space of many particles follows the conventional procedure.
Therefore by introducing the notation $x^\mu=(t,x^1,x^2)=(t,{\bf x})$ and $k_\mu=(\omega_k,k_1,k_2)=(\omega_k,{\bf k})$, we can
express a scalar field as follows

\begin{equation}
\phi({\bf x},t)=\int \frac{d^2 k}{\left(2\pi\right)^2 2\omega_k}
\left[a({\bf k}){}_{\theta\sigma}\langle a\ b|e^{i{\bf \hat k}{\bf \hat x}-i\omega_k t}|a\ b\rangle_{\theta\sigma}
+a^{\dagger}({\bf k}){}_{\theta\sigma}\langle a\ b|e^{-i{\bf \hat k}{\bf \hat x}
+i\omega_k t}|a\ b\rangle_{\theta\sigma}\right]
\label{scalar_field}
\end{equation}
whose quantization is achieved in the usual way by postulating

\begin{equation}
\left[a({\bf k}),a^{\dagger}({\bf k})\right]=\delta^2\left({\bf k}-{\bf k^{\prime}}\right).
\end{equation}
Again the symbol ``$\ \hat{}\ $'' in (\ref{scalar_field}) indicates operators acting on the Fock space ${\cal H}_\theta\cong {\cal H}_\sigma$.
If the expectation values are calculated, we get

\begin{equation}
\phi({\bf x},t)=\int \frac{d^2 k}{\left(2\pi\right)^2 2\omega_k}
\left[a({\bf k})e^{i{\bf kx}-i\omega_k t}+a^{\dagger}({\bf k})e^{-i{\bf kx}+i\omega_k t}\right]
\exp\left[-\frac{1}{4}\theta{\bf k}^2\right]\exp\left[-\frac{1}{4}\sigma{\bf x}^2\right].
\end{equation}
The propagator is obtained in the usual way by considering the expectation value of the time
ordered product of the quantum field operator at two different positions with respect to 
the vacuum $|0 \rangle$ state of the quantum field

\begin{equation}
G(x-x^\prime)=i\Delta_F(x-x^\prime)=\langle 0|T[\phi(x)\phi(x^\prime)]|0 \rangle,
\label{expression_propagator}
\end{equation}
where $T$ denotes the time ordering operator.
Without loss of generality the coordinate system can always be chosen in such a way that one of the coordinates between $x$ and
$x^{\prime}$ lies at the origin. As a result, the expression of the propagator only depends on the difference of the variables
$z=x-x^{\prime}$ if the coordinates are accordingly chosen as $x=z$ and $x^{\prime}=0$.
Thus the propagator can be formulated in such a way that it only depends on the difference of the coordinates

\begin{eqnarray}
\langle 0|T[\phi(z)\phi(0)]|0 \rangle
=\int \frac{d^2 k}{(2\pi)^2 2 \omega_k}
\left[e^{i{\bf kz}-i\omega_k t_z}\theta(t_z)+e^{-i{\bf kz}+i\omega_k t_z}\theta(-t_z)\right]
\exp\left[-\frac{1}{2}\theta{\bf k}^2\right]\exp\left[-\frac{1}{4}\sigma{\bf z}^2\right].
\label{expression_operator_product2}
\end{eqnarray}
From now on, we can follow the usual procedure of the derivation of a propagator. By using

\begin{equation}
\theta(t)e^{-i\omega_k t}=\frac{1}{2\pi i}\int^{\infty}_{-\infty}dE \frac{e^{-iEt}}{\omega_k-E-i\epsilon},
\label{step_function}
\end{equation}
equation ($\ref{expression_operator_product2}$) can be transformed to

\begin{equation}
\langle 0|T[\phi(z)\phi(0)]|0 \rangle=G(z)=i\int \frac{d^3 k}{(2\pi)^3}
\frac{e^{-ikz}}{k^2-m^2+i\epsilon}\exp\left[-\frac{1}{2}\theta{\bf k}^2\right]\exp\left[-\frac{1}{4}\sigma{\bf z}^2\right]
\label{expression_operator_product3}
\end{equation}
which is the propagator for a scalar field in the case of a (2+1) dimensional spacetime with the special
manifestation ($\ref{algebra_two_dimensions}$) and ($\ref{algebra_two_dimensions_matrices}$) 
of ($\ref{Algebra}$). The propagator in momentum space is obtained by a Fourier transformation as usual,
since the above expression is already expressed in terms of expectation values. 
As a result, the propagator in momentum space reads

\begin{eqnarray}
G(k)&=&\int \frac{d^2 k^{\prime}}{(2\pi)^2}\frac{\rho_\sigma({\bf k,k^{\prime}})\ e^{-\frac{1}{2}\theta{\bf k}^{\prime 2}}}
{\omega_k^2-{\bf k^{\prime 2}}-m^2+i\epsilon},
\label{propagator_momentum_space_integrated}
\end{eqnarray}
where

\begin{eqnarray}
\rho_\sigma({\bf k,k^{\prime}})&=&\frac{1}{\pi\sigma}\ e^{-\frac{(\bf k-k^{\prime})^2}{\sigma}}.
\end{eqnarray}
We notice that the IR behavior of the propagator is now controlled by the function $\rho_\sigma({\bf k,k^{\prime}})$,
which in the limit $\sigma\to 0$ becomes a delta function, leading to the result

\begin{equation}
G(k)=\frac{ e^{-\frac{1}{2}\theta{\bf k}^{2}}}
{\omega_k^2-{\bf k^{ 2}}-m^2+i\epsilon}\\
\end{equation}
for the UV  finite propagator \cite{NCQFT}. Conversely, for $\sigma\neq 0$, ($\ref{propagator_momentum_space_integrated}$)
provides a Gaussian modulation of the propagator, preventing the resolution of momenta smaller that $\sqrt\sigma$.
In other words the distribution of $\bf k^\prime$ has a bell shape of width $\sqrt\sigma$ and it is no longer peaked on the
value $\bf k$ due to the intrinsic uncertainty we have introduced in the formulation.

\section{The case of a (3+1)-dimensional spacetime}

The ($2+1$)-dimensional case can be extended to the more realistic case of a ($3+1$)-dimensional Minkowski spacetime.
Let us start from the algebra ($\ref{Algebra}$). As a first remark, we proceed along the line of \cite{NCQFT}
recalling that our matrices describing the noncommutativity are Lorentz tensors.
For later convenience, we are initially working with an Euclidean signature without loss of generality.
Then one can exploit the property for which any antisymmetric tensor can be written in a block diagonal form as

\begin{equation}
\Theta=\left(\begin{matrix}0&\theta_1&0&0\\-\theta_1&0&0&0\\0&0&0&\theta_2\\0&0&-\theta_2&0\end{matrix}\right)\quad,\quad
\Sigma=\left(\begin{matrix}0&-\sigma_1&0&0\\\sigma_1&0&0&0\\0&0&0&-\sigma_2\\0&0&\sigma_2&0\end{matrix}\right).
\end{equation}
In a more compact form this can be written as

\begin{equation}
\Theta=\mathrm{diag}\left(\ \hat{\theta}^1,\  \hat{\theta}^2\ \right)\quad,\quad
\Sigma=\mathrm{diag}\left(\ \hat{\sigma}_1,\  \hat{\sigma}_2\ \right),
\end{equation}
where

\begin{equation}
\hat\theta^{i}=\theta^i\left( \begin{array}{cc}
0 & 1  \\
-1 & 0 \\
\end{array} \right)\quad,\quad
\hat\sigma_{j}=-\sigma_j\left( \begin{array}{cc}
0 & 1  \\
-1 & 0 \\
\end{array} \right)
\end{equation}
with $i,j=1,2$. As a result the noncommutative relations actually foliate the spacetime in two planes. Lorentz covariance
implies that different observers get different noncommutativity  due to different projections of noncommutative plane
in their reference frames. Conversely what is invariant, namely a fact recognized by any observer, is the presence of
these two planes which foliate the spacetime. Thus the problem is reduced to an effective two-dimensional problem
\begin{equation}
\hat A_i=\frac{1}{\sqrt{2\theta_i}}\left(\hat Q_i^1+i\hat Q_i^2\right) \quad,\quad \hat A^{\dagger}_i
=\frac{1}{\sqrt{2\theta_i}}\left(\hat Q^1_i-i\hat Q_i^2\right)\quad,\quad
\hat B_j=\frac{1}{\sqrt{2\theta_j}}\left(\hat P_{j2}+i\hat P_{j1}\right) \quad,\quad \hat B^{\dagger}_j
=\frac{1}{\sqrt{2\theta_j}}\left(\hat P_{j2}-i\hat P_{j1}\right),
\end{equation}
where $(\hat Q_i^1,\hat Q_i^2)$ are coordinate operators in the $i$-th plane, while $(\hat P_{j1},\hat P_{j2})$ are
the corresponding ones in momentum space. These operators fulfil the following commutation relations

\begin{eqnarray}
\left[\hat A_i,\hat A_j^{\dagger}\right]=\delta_{ij}
\quad,\quad \left[\hat B_i,\hat B_j^{\dagger}\right]=\delta_{ij}.
\end{eqnarray}
Coherent states can be defined as 

\begin{equation}
|a\rangle_{\theta} =\prod_i e^{-{\frac{|a_i|^2}{2}}}\sum_{n=0}^{\infty}{\frac{a_i^n}{\sqrt{n!}}}|n\rangle_{\theta_i}
=\prod_i e^{-{\frac{|a_i|^2}{2} }}e^{a_i\hat A_i^\dagger}|0\rangle_{\theta_i},
\end{equation}
with

\begin{equation}
\hat A_i|a\rangle_{\theta}=a_i|a\rangle_{\theta}
\end{equation}
and

\begin{equation}
|b\rangle_{\sigma}=\prod_j e^{-{\frac{|b_j|^2}{2} }}\sum_{n=0}^{\infty}{\frac{b_j^n}{\sqrt{n!}}}|n\rangle_{\sigma_j}
=\prod_j e^{-{\frac{|b_j|^2}{2}}}e^{b_j\hat B_j^\dagger}|0\rangle_{\sigma_j},
\end{equation}
with
\begin{equation}
\hat B_i|b\rangle_\sigma=b_i|b\rangle_\sigma.
\end{equation}
With the help of these states, we can define expectation values for the positions and momenta according to 

\begin{eqnarray}
{}_\theta\langle a|\hat Q_i^1 |a\rangle_\theta=Q_i^1\quad,\quad
{}_\theta\langle a|\hat Q_i^2 |a\rangle_\theta=Q_i^2\quad,\quad
{}_\sigma\langle b|\hat P_{j1} |b\rangle_\sigma=P_{j1}\quad,\quad
{}_\sigma\langle b|\hat P_{j2} |b\rangle_\sigma=P_{j2}.
\label{expectation_values_3+1}
\end{eqnarray}
As a result, the expectation values of the plane wave have the following shape

\begin{eqnarray}
{}_{\theta\sigma}\langle a \ b |\exp\left[i\sum_i\mathcal{Q}_i\cdot \mathcal{P}_i \right]|a\ b\rangle_{\theta\sigma}
=\exp\left[i\sum_i \vec P_i\cdot \vec Q_i\right]\exp\left[-\frac{1}{4}\sum_i\left(\theta_i \vec P_i^2+\sigma_i \vec Q_i^2 \right)\right].
\label{modified_plane_waves_3+1}
\end{eqnarray}
The above expression clearly depends on the orientations of the noncommutative plane due to the presence of different
parameters $\theta_i$ and $\sigma_i$. To restore a Lorentz covariant formalism, we follow the lines in \cite{NCQFT} and we set
$\theta_i=\theta$ and $\sigma_i=\sigma$. Given this point and in analogy to the (2+1)-dimensional scenario we are ready to
calculate the scalar field propagator. We will use the following notation for the position
\ $x^\mu=(t,x^1,x^2,x^3)=(t,{\bf x})=(Q^1_1,Q^2_1,Q^1_2,Q^2_2)$ and the momentum
\ $k_\mu=(k_0,k_1,k_2,k_3)=(k_0,{\bf k})=(P_{11},P_{12},P_{21},P_{22})$.
Starting from the scalar field written as

\begin{equation}
\phi({\bf x},t)=\int \frac{d^3 k}{\left(2\pi\right)^3 2 k_0}
\left[a({\bf k}){}_{\theta\sigma}\langle a\ b |e^{i\hat k_\mu \hat x^\mu}|a\ b\rangle_{\theta\sigma}+a^{\dagger}({\bf k})
{}_{\theta\sigma}\langle a\ b| e^{-i\hat k_\mu \hat x^\mu}|a\ b \rangle_{\theta\sigma}\right],
\end{equation}
with

\begin{equation}
\left[a({\bf k}),a^{\dagger}({\bf k})\right]=\delta^3\left({\bf k}-{\bf k^{\prime}}\right).
\end{equation}
we get

\begin{equation}
\phi({\bf x},t)=\int \frac{d^3 k}{\left(2\pi\right)^3 2 k_0}
\left[a({\bf k})e^{ik_\mu x^\mu}+a^{\dagger}({\bf k})
e^{-i k_\mu x^\mu}\right]\exp\left[-\frac{1}{4}\theta k_\mu k^\mu\right]\exp\left[-\frac{1}{4}\sigma x_\mu x^\mu\right].
\label{scalar_field_3+1}
\end{equation}
The related expression of the propagator reads

\begin{eqnarray}
G(z)&=&\langle 0|T[\phi(z)\phi(0)]|0 \rangle\\&=&\int \frac{d^3 k}{(2\pi)^3 2 k_0}
\left[e^{i k_\mu z^\mu}\theta(z_0)+e^{-i k_\mu z^\mu}\theta(-z_0)\right]
\exp\left[-\frac{1}{2}\theta k_\mu k^\mu\right]\exp\left[-\frac{1}{4}\sigma z_\mu z^\mu\right]\nonumber
\label{expression_operator_product_3+1}
\end{eqnarray}
where, without loss of generality, we have again chosen the coordinate system in such a way that one of the coordinates
lies at the origin. This means that we assume that $x=z$ and $x^\prime=0$ and therefore the propagator turns out to depend 
only on the variable $z=x-x^\prime$. By using (\ref{step_function}) and following the usual procedure, one can write 

\begin{eqnarray}
G(z)=\langle 0|T[\phi(z)\phi(0)]|0 \rangle=i\int \frac{d^4 k}{(2\pi)^4}\frac{e^{-ikz}}{k^2+m^2+i\epsilon}
\exp\left[-\frac{1}{2}\theta k^2-\frac{1}{4}\sigma z^2\right].
\label{propagator_position_space_3+1}
\end{eqnarray}
As a result the corresponding propagator in momentum space reads 

\begin{eqnarray}
G(k)=\int \frac{d^4 k^{\prime}}{(2\pi)^4}
\frac{\rho_\sigma (k,k^\prime)}{k^{\prime 2}+m^2+i\epsilon}
e^{-\frac{1}{2}\theta k^{\prime 2}},
\label{finitepropag}
\end{eqnarray}
where the modulation function

\begin{equation}
\rho_\sigma(k,k^\prime)=\frac{1}{\pi^2\sigma^2}e^{-\frac{(k-k^{\prime})^2}{\sigma}}
\end{equation}
accounts for the uncertainty of the momentum.
In the limit $\sigma\to 0$, there is no longer any uncertainty of momentum, which is exactly peaked on the value $k$.
The modulation function $\rho_\sigma(k,k^\prime)$ becomes a Dirac delta $\rho_\sigma(k,k^\prime)\to\delta(k,k^\prime)$
and the propagator reads

\begin{equation}
G(k)=\frac{e^{-\frac{1}{2}\theta k^{2}}}{k^{2}+m^2+i\epsilon},
\end{equation}
reproducing the UV finite propagator as in \cite{NCQFT}. Conversely for $\sigma\neq 0$, 
the modulation function $\rho(k,k^\prime)$ prevents better resolutions in momentum space than $\sqrt\sigma$.
The integral (\ref{finitepropag}) is actually endowed with a IR cutoff. As a result, if we consider the limit
$m\to 0$, the propagator converges even for vanishing values of $k$. As shown in the appendix we have

\begin{equation}
G(0)\approx\frac{1}{\sigma}
\end{equation}
for $m\to 0$. As a conclusion, the propagator is finite both in the UV and IR regime.
In support of this claim, we also provide  the following integral representation for the Green's function

\begin{equation}
G(x-x')=\frac{e^{\theta m^2/2}}{16\pi^{2}} \int_{\theta/(2+\sigma\theta)}^{1/\sigma}\frac{ds}{s^2}
\ e^{-m^2s/(1-\sigma s)}\ e^{-(x-x')^2/4s}
\label{integral}
\end{equation}
that after manipulations can be cast in the more familiar form

\begin{equation}
G(x-x')=\int_0^\infty ds \ e^{-m^2s}\ \frac{e^{-(\sigma+\frac{1}{s+\theta/2})(x-x')^2/4}}{[4\pi(s+\theta/2)]^{2}}.
\end{equation}
In the massless case, we obtain after integration

\begin{equation}
G(x-x')=\frac{e^{-\frac{1}{4}\sigma (x-x')^2}}{4\pi^{2}(x-x')^{2}} \ \left(1-e^{-(x-x')^2/2\theta}\right).
\end{equation}
In addition, if the coordinates are Wick rotated back to Minkowski signature, we can express the momentum space
propagator as \footnote{One must have care of the correct definition of the damping terms within the chosen signature.}

\begin{eqnarray}
G(k)=\frac{i}{\pi^2\sigma^2}\int \frac{d^4 k^{\prime}}{(2\pi)^4}
\frac{e^{-\frac{(k-k^{\prime})^2}{\sigma}}}{k^{\prime 2}-m^2+i\epsilon}
e^{-\frac{1}{2}\theta k^{\prime 2}}.
\end{eqnarray}
For completeness we study also the case of a fermionic field. Since the deformation we are introducing concerns the shape
of the Fourier modes only, a fermionic field can be treated in analogy to a scalar field ($\ref{scalar_field_3+1}$) and reads 

\begin{eqnarray}
\psi({\bf x},x_0)=\sum_{\pm s}\int \frac{d^3 k}{\left(2\pi\right)^3}\sqrt{\frac{m}{k_0}}
\left[b_s({\bf k})u_s({\bf k})e^{i k_\mu x^\mu}
+d^{\dagger}_s({\bf k})v_s({\bf k})e^{-i k_\mu x^\mu}\right]
\exp\left[-\frac{1}{4}\theta k_\mu k^\mu\right]\exp\left[-\frac{1}{4}\sigma x_\mu x^\mu\right],\nonumber\\
\psi^{\dagger}({\bf x},x_0)
=\sum_{\pm s}\int \frac{d^3 k}{\left(2\pi\right)^3}\sqrt{\frac{m}{k_0}}
\left[b^{\dagger}_s({\bf k}) u^{\dagger}_s({\bf k}) e^{-i k_\mu x^\mu}
+d_s({\bf k}) v^{\dagger}_s({\bf k}) e^{i k_\mu x^\mu}\right]
\exp\left[-\frac{1}{4}\theta k_\mu k^\mu\right]\exp\left[-\frac{1}{4}\sigma x_\mu x^\mu\right],\nonumber\\
\label{fermion_field_3+1}
\end{eqnarray}
where the coefficients of the quantized fields fulfil as usual the anti-commutation relations

\begin{equation}
\{b_s \left({\bf k}\right), b_{s^{\prime}}^{\dagger}\left(\bf k^{\prime}\right)\}
=\delta^3\left({\bf k-\bf k^{\prime}}\right)\delta_{s,s^{\prime}}\quad,\quad
\{d_s \left({\bf k}\right), d_{s^{\prime}}^{\dagger}\left(\bf k^{\prime}\right)\}
=\delta^3\left({\bf k-\bf k^{\prime}}\right)\delta_{s,s^{\prime}}.
\end{equation}
As usual there is a factor $\left[k\!\!\!/+m\right]$ with respect the calculation for the scalar field

\begin{eqnarray}
G(z)=\langle 0|T[\psi(z)\bar \psi(0)]|0 \rangle
=i\int \frac{d^4 k}{(2\pi)^3}\frac{e^{-ikz}\left[k\!\!\!/+m\right]}{k^2+m^2+i\epsilon}
\exp\left[-\frac{1}{2}\theta k^2-\frac{1}{4}\sigma z^2\right],
\label{propagator_position_space}
\end{eqnarray}
where $\bar \psi(x)=\psi^{\dagger}(x)\gamma^0 $ and $k\!\!\!/=\gamma^\mu k_\mu$.
The corresponding propagator in momentum space reads 

\begin{eqnarray}
G(k)=\frac{i}{\pi^2\sigma^2}\int \frac{d^4 k^{\prime}}{(2\pi)^4}
\frac{e^{-\frac{(k-k^{\prime})^2}{\sigma}}}{k\!\!\!/^{\prime}+m+i\epsilon}
e^{-\frac{1}{2}\theta k^{\prime 2}}.
\end{eqnarray}
If the coordinates are now Wick rotated back, the propagator in Minkowski space reads

\begin{equation}
G(k)=\frac{i}{\pi^2\sigma^2}\int \frac{d^4 k^{\prime}}{(2\pi)^4}
\frac{e^{-\frac{(k-k^{\prime})^2}{\sigma}}}{k\!\!\!/^{\prime}-m+i\epsilon}
e^{-\frac{1}{2}\theta k^{\prime 2}}.
\label{Minkowski_propagator_momentum_space}
\end{equation}
Here the IR finiteness can be shown in an analogous way of the scalar field case.
Since momentum vectors of particles are time-like or light-like,
assuming  the signature (1,-1,-1,-1) for the Minkowski metric, we maintain the damping
behaviour of our new factors within the integral of the
expression for the propagator in momentum space ($\ref{Minkowski_propagator_momentum_space}$).
However even if it is nice to display formulas in Minkowski space, our general prescription is to perform virtual momentum integrations in Euclidean spacetime, where the exponential terms have unambiguous sign. Only the
final results are, then, continued into Minkowski spacetime and the physical implications
are discussed.

\section{Summary and Discussion}

In this paper, we considered an extension of the usual Heisenberg algebra, assuming nontrivial commutation relations for
position as well as for momentum operators. As a result we have introduced a minimal length and a minimum momentum as the
smallest scales of resolution in coordinate and momentum space. To define suitable coordinates and momenta, we have extended
the approach in \cite{NCQFT}, calculating expectation values of operators on coherent states referring to two Fock spaces,
one built starting from coordinate operators and the other from momentum operators. As a result we obtained that conventional
plane waves are modulated by two factors, $e^{-\frac{1}{4}\theta p^2}$ and $e^{-\frac{1}{4}\sigma x^2}$, corresponding to
damping terms in position and momentum space. This result is in agreement with the one which was found in \cite{GUPM}. As a
remark we stress that these modulations are physically consistent, since the modified plane waves become square integrable
functions and reliably describe physical states. After this investigation, we have considered the consequences for quantum
field theory, which now is endowed with two cutoffs, an UV cutoff and an IR one. We derived the propagator, which shows
regularity at higher momenta and at lower momenta in the massless case, too. Finally we showed that this property holds both,
for the scalar and the fermionic field. In the case of  a (3+1)-dimensional space-time, Lorentz invariance of the plane waves can be maintained only, if the noncommutativity parameters
referring to the two planes are assumed to be equal: $\theta_1=\theta_2$ and $\sigma_1=\sigma_2$.
With respect to this, we think that the appearance of UV as well as IR cutoffs will deserve further investigations to make
compelling predictions for current research activities in phenomenological and experimental frameworks.
For instance, IR Effects on Quantum Mechanics are currently under investigation \cite{Mirza:2009mh}. It would also be very interesting to study within our approach the phenomenological
consequences which are so far just explored in other formulations of noncommutative geometry like that based on the use of the star product \cite{PhenomenologicalConsequencesNCG}.\\
In contrast to the star product approach to noncommutative geometry
the coherent state approach implies no modification of the interaction
structure between fields. Therefore it changes the kinematics of quantum
mechanics and thus of quantum field theory, but not the dynamics. This
is related to the fact that the extended algebra has to be interpreted
as a modification of the properties of the operators $\hat q^i$ and $\hat p^i$
referring to the state of a particle. In such a way we can still define in a quasi-classical way the concept of coordinate and momentum. On the contrary, in the star product approach the
noncommutative coordinates modify the structure of the product among functions, a fact that has
 an influence on interaction terms (see  \cite{ChangingInteractionStructureNCG} for
example). Since this is not the case in the coherent state approach, we have that
 within the Feynman rules just the propagator changes but the vertices
remain completely unchanged. Because of this, we conclude that no UV/IR mixing occurs in the sense
of \cite{Minwalla:1999px} according to our scenario. This conclusion is supported by analogous arguments in \cite{NCQFT} for the specific case of noncommutative coordinates and commuting momenta.

\textit{Acknowledgement}: M. K. would like to thank the Messer Stiftung for financial support. P.N. is supported by the
Helmholtz International Center for FAIR within the framework of the LOEWE program (Landesoffensive zur Entwicklung
Wissenschaftlich-\"{O}konomischer Exzellenz) launched by the State of Hesse. 

\appendix*
\section{}
\noindent
We start from the propagator (\ref{finitepropag}), to show its IR regularity. We assume without loss of generality
that the UV parameter is vanishing $\theta=0$ and we end up with 

\begin{equation}
G(k)=\frac{1}{\pi^2\sigma^2}\int \frac{d^4k^\prime}{(2\pi)^4}\frac{e^{(k-k^\prime)^2/\sigma}}{(k^\prime)^2+m^2}.
\end{equation}
We want to show that $G(k)$ is finite in the limit $m\to 0$ for $k=0$. Thus we write 

\begin{equation}
G(0)=\frac{1}{\pi^2\sigma^2}\int \frac{d^4k^\prime}{(2\pi)^4}\frac{e^{-(k^\prime)^2/\sigma}}{(k^\prime)^2+m^2},
\end{equation}
which reads

\begin{equation}
G(0)=\frac{1}{\pi^2\sigma^2}\int_{-\infty}^\infty \frac{d(k^0)^\prime}{2\pi} e^{-[(k^0)^\prime]^2/\sigma}
\int_0^\infty \frac{d|\vec k^\prime||\vec k^\prime|^2}{2\pi^2}\frac{e^{-(k^\prime)^2/\sigma}}{|\vec k^\prime|^2+[(k^0)^\prime]^2+m^2},
\end{equation}
where we have used the Euclidean relation $(k^\prime)^2=|\vec k^\prime|^2+[(k^0)^\prime]^2$. 
Now calling $\alpha^2\equiv [(k^0)^\prime]^2+m^2$, $y\equiv (k^0)^\prime/\sqrt\sigma$
and $\sigma x\equiv |\vec k^\prime|^2+ \alpha^2$ we end up with

\begin{equation}
G(0)=\frac{1}{2\pi^2\sigma}\int_{-\infty}^\infty \frac{dy}{2\pi} e^{\alpha^2/\sigma} e^{-y^2}
\int_{\alpha^2/\sigma}^\infty \frac{dx}{2\pi^2}( x-\alpha^2/\sigma)^{1/2}\frac{e^{-x}}{x}.\label{int}
\end{equation}
For $\sigma\neq 0$, the above integral is defined for every $\alpha$. More specifically in the limit
$m\to 0$ we have that $\alpha^2=[(k^0)^\prime]^2$ and we can define a function $F$

\begin{equation}
F(y)=\int_{y^2}^\infty dx ( x-y^2)^{1/2}\frac{e^{-x}}{x}
\end{equation} 
which goes to $0$ as $y^2\to \infty$, while it is a constant for $y^2\to 0$,  i.e.

\begin{equation}
F(0)=\int_{0}^\infty dxx^{-1/2}e^{-x}=\Gamma(1/2)=\sqrt{\pi}.
\end{equation}
Since it is difficult to estimate the integral (\ref{int}) directly, we make use of the following inequality

\begin{equation}
F(y)\leq\int_{y^2}^\infty dx x^{-1/2}e^{-x}\equiv \Gamma(1/2 ;y^2)
\end{equation}
with $\Gamma(1/2 ;y^2)$ the incomplete upper Gamma function. 
Therefore the integral (\ref{int}) turns out to be limited

\begin{equation}
G(0)\leq\frac{1}{4\pi^5\sigma}\int_{-\infty}^\infty dy \Gamma(1/2 ;y^2)
\end{equation}
 and being

\begin{equation}
\int_{-\infty}^\infty dy \Gamma(1/2 ;y^2)=2
\end{equation}
we conclude that

\begin{equation}
G(0)\leq\frac{1}{2\pi^5\sigma},
\end{equation}
showing the IR cutoff $1/\sigma$ at work in curing IR divergences of the propagator.


\begin{thebibliography}{99}

\bibitem{ReviewQG}
  J.~H.~Schwarz and N.~Seiberg,
  Rev.\ Mod.\ Phys.\  {\bf 71} (1999) S112
  [arXiv:hep-th/9803179].

  H.~Nicolai, K.~Peeters and M.~Zamaklar,
  Class.\ Quant.\ Grav.\  {\bf 22} (2005) R193
  [arXiv:hep-th/0501114].

  S.~Hossenfelder,
  Mod.\ Phys.\ Lett.\  A {\bf 19} (2004) 2727
  [arXiv:hep-ph/0410122].

  C.~Kiefer,
  Annalen Phys.\  {\bf 15} (2005) 129
  [arXiv:gr-qc/0508120].


\bibitem{ReviewNCG}
  G.~Landi,
  arXiv:hep-th/9701078.
  
  M.~R.~Douglas and N.~A.~Nekrasov,
  Rev.\ Mod.\ Phys.\  {\bf 73}, 977 (2001)
  [arXiv:hep-th/0106048].
  
  R.~J.~Szabo,
  Phys.\ Rept.\  {\bf 378} (2003) 207
  [arXiv:hep-th/0109162].


\bibitem{DeWitt}B. DeWitt, {\em Gravitation}, edited by L. Witten (Wiley, New York), pp. 266-381.


\bibitem{Snyder:1946qz}
  H.~S.~Snyder,
  Phys.\ Rev.\  {\bf 71} (1947) 38.


\bibitem{Seiberg:1999vs}
  N.~Seiberg and E.~Witten,
  JHEP {\bf 9909}, 032 (1999)
  [arXiv:hep-th/9908142].


\bibitem{GUP}
  A.~Kempf,
  J.\ Math.\ Phys.\  {\bf 35}, 4483 (1994);
  [arXiv:hep-th/9311147].

  M.~Maggiore,
  Phys.\ Lett.\  B {\bf 319}, 83 (1993);
  [arXiv:hep-th/9309034].

  M.~Maggiore,
  Phys.\ Rev.\  D {\bf 49}, 5182 (1994);
  [arXiv:hep-th/9305163].

  M.~Maggiore,
  Phys.\ Lett.\  B {\bf 304}, 65 (1993);
  [arXiv:hep-th/9301067].

  A.~Kempf, G.~Mangano and R.~B.~Mann,
  Phys.\ Rev.\  D {\bf 52}, 1108 (1995).
  [arXiv:hep-th/9412167].


\bibitem{NCBHs}
  P.~Nicolini,
  J. Phys. A  {\bf 38}, L631 (2005)
  [arXiv:hep-th/0507266];
 
  P.~Nicolini, A.~Smailagic and E.~Spallucci,
  Phys. Lett.  B {\bf 632}, 547 (2006)
  [arXiv:gr-qc/0510112];

  P.~Nicolini, A.~Smailagic and E.~Spallucci,
  arXiv:hep-th/0507226;

  S.~Ansoldi, P.~Nicolini, A.~Smailagic and E.~Spallucci,
  Phys. Lett.  B {\bf 645}, 261 (2007)
  [arXiv:gr-qc/0612035];


  E.~Spallucci, A.~Smailagic and P.~Nicolini,
  Phys. Lett.  B {\bf 670}, 449 (2009)
  [arXiv:0801.3519 [hep-th]];
  
 
  

  P.~Nicolini and E.~Spallucci,
  Class.\ Quant.\ Grav.\  {\bf 27}, 015010 (2010)
  [arXiv:0902.4654 [gr-qc]];
 

  D.~Batic and P.~Nicolini,
  Phys.\ Lett.\  B {\bf 692}, 32 (2010).

  
  
  A.~Smailagic and E.~Spallucci,
  Phys.\ Lett.\  B {\bf 688}, 82 (2010)
  [arXiv:1003.3918 [hep-th]].

  
\bibitem{review}
  P.~Nicolini,
  Int. J. Mod. Phys.  A {\bf 24}, 1229 (2009)
  [arXiv:0807.1939 [hep-th]];


\bibitem{NCcosmo}
  M.~Rinaldi,
  arXiv:0908.1949 [gr-qc];

  K.~Nozari and S.~Akhshabi,
  Phys.\ Lett.\  B {\bf 683}, 186 (2010)
  [arXiv:0911.4418 [hep-th]].
  

\bibitem{NCthermo} 
  
  R.~Banerjee, B.~R.~Majhi and S.~Samanta,
  Phys.\ Rev.\  D {\bf 77}, 124035 (2008)
  [arXiv:0801.3583 [hep-th]].

  W.~H.~Huang and K.~W.~Huang,
  Phys.\ Lett.\  B {\bf 670}, 416 (2009)
  [arXiv:0808.0324 [hep-th]];

  R.~Casadio and P.~Nicolini,
  JHEP {\bf 0811}, 072 (2008)
  [arXiv:0809.2471 [hep-th]];
  

   P.~Nicolini,
  Phys.\ Rev.\  D {\bf 82}, 044030 (2010).


\bibitem{NCwormhole}
  R.~Garattini and F.~S.~N.~Lobo,
  Phys.\ Lett.\  B {\bf 671}, 146 (2009)
  [arXiv:0811.0919 [gr-qc]];


\bibitem{NCApplications}
  R.~Casadio, P.~H.~Cox, B.~Harms and O.~Micu,
  Phys. Rev.  D {\bf 73}, 044019 (2006)
  [arXiv:gr-qc/0510115];

  R.~Casadio, A.~Gruppuso, B.~Harms and O.~Micu,
  Phys. Rev.  D {\bf 76}, 025016 (2007)
  [arXiv:0704.2251 [hep-th]];

  N.~Nicolaevici,
  Phys. Rev.  D {\bf 78}, 088501 (2008);

  P.~Nicolini and M.~Rinaldi,
  arXiv:0910.2860 [hep-th];

\bibitem{NCparticle}
  M.~Rinaldi,
 Mod.\ Phys.\ Lett.\  A {\bf 25}, 2805 (2010)
  arXiv:1003.2408 [hep-th].


\bibitem{NCpheno}
  T.~G.~Rizzo,
  JHEP {\bf 0609}, 021 (2006)
  [arXiv:hep-ph/0606051];



  D.~M.~Gingrich,
  JHEP {\bf 1005}, 022 (2010).

  

\bibitem{NCspectral}
 L.~Modesto and P.~Nicolini,
  Phys.\ Rev.\  D {\bf 81}, 104040 (2010).

 P.~Nicolini and E.~Spallucci,
  arXiv:1005.1509 [hep-th].


\bibitem{Infeld}  
  L.~Infeld and A.~E.~Schild,
  Phys.\ Rev.\  {\bf 68}, 250 (1945);

  L.~Infeld and A.~E.~Schild,
  Phys.\ Rev.\  {\bf 70}, 410 (1946).


\bibitem{Graviton}
  C.~Deffayet, G.~R.~Dvali and G.~Gabadadze,
  Phys.\ Rev.\  D {\bf 65}, 044023 (2002)
  [arXiv:astro-ph/0105068].

  N.~Arkani-Hamed, H.~C.~Cheng, M.~A.~Luty and S.~Mukohyama,
  JHEP {\bf 0405}, 074 (2004)
  [arXiv:hep-th/0312099].

  G.~Dvali, S.~Hofmann and J.~Khoury,
  Phys.\ Rev.\  D {\bf 76}, 084006 (2007)
  [arXiv:hep-th/0703027].


\bibitem{HF}
  B.~S.~DeWitt,
  Phys.\ Rept.\  {\bf 19}, 295 (1975).

  P.~R.~Anderson, W.~A.~Hiscock and D.~A.~Samuel,
  Phys.\ Rev.\ Lett.\  {\bf 70}, 1739 (1993).

  P.~R.~Anderson, W.~A.~Hiscock and D.~A.~Samuel,
  Phys.\ Rev.\  D {\bf 51}, 4337 (1995).

  V.~P.~Frolov, P.~Sutton and A.~Zelnikov,
  Phys.\ Rev.\  D {\bf 61}, 024021 (2000)
  [arXiv:hep-th/9909086].

  R.~Balbinot, A.~Fabbri, V.~P.~Frolov, P.~Nicolini, P.~Sutton and A.~Zelnikov,
  Phys.\ Rev.\  D {\bf 63}, 084029 (2001)
  [arXiv:hep-th/0012048].

  R.~Balbinot, A.~Fabbri, P.~Nicolini and P.~J.~Sutton,
  Phys.\ Rev.\  D {\bf 66}, 024014 (2002)
  [arXiv:hep-th/0202036].


\bibitem{USEP}
  F.~Piazza,
  arXiv:0904.4299 [hep-th].

  F.~Piazza,
  New J.\ Phys.\  {\bf 11}, 113050 (2009)
  [arXiv:0907.0765 [hep-th]].

  S.~Nesseris, F.~Piazza and S.~Tsujikawa,
  arXiv:0910.3949 [astro-ph.CO].


\bibitem{cosmoenergyfluc}  
  S.~W.~Hawking,
  Phys.\ Lett.\  B {\bf 115}, 295 (1982).

  A.~H.~Guth and S.~Y.~Pi,
  Phys.\ Rev.\ Lett.\  {\bf 49}, 1110 (1982).

  J.~M.~Bardeen, P.~J.~Steinhardt and M.~S.~Turner,
  Phys.\ Rev.\  D {\bf 28}, 679 (1983).

  T.~Padmanabhan,
  Phys.\ Rev.\ Lett.\  {\bf 60}, 2229 (1988).

  T.~Padmanabhan, T.~R.~Seshadri and T.~P.~Singh,
  Phys.\ Rev.\  D {\bf 39}, 2100 (1989).


\bibitem{Padmanabhan:2004qc}
  T.~Padmanabhan,
  Class.\ Quant.\ Grav.\  {\bf 22}, L107 (2005)
  [arXiv:hep-th/0406060].

  B.~Guberina, R.~Horvat and H.~Nikolic,
  Phys.\ Rev.\  D {\bf 72}, 125011 (2005)
  [arXiv:astro-ph/0507666].

  A.~Sheykhi,
  Class.\ Quant.\ Grav.\  {\bf 27}, 025007 (2010)
  [arXiv:0910.0510 [hep-th]].


\bibitem{NCQFT}
  A.~Smailagic and E.~Spallucci,
  J. Phys. A  {\bf 36}, L467 (2003)
  [arXiv:hep-th/0307217];
 
  A.~Smailagic and E.~Spallucci,
  J. Phys. A  {\bf 36}, L517 (2003)
  [arXiv:hep-th/0308193];

  A.~Smailagic and E.~Spallucci,
  J. Phys. A  {\bf 37}, 1 (2004)
  [Erratum-ibid.  A {\bf 37}, 7169 (2004)]
  [arXiv:hep-th/0406174];

  E.~Spallucci, A.~Smailagic and P.~Nicolini,
  Phys. Rev.  D {\bf 73}, 084004 (2006)
  [arXiv:hep-th/0604094];

  
\bibitem{NCGEOM}
  S.~Cho, R.~Hinterding, J.~Madore and H.~Steinacker,
  Int.\ J.\ Mod.\ Phys.\  D {\bf 9}, 161 (2000)
  [arXiv:hep-th/9903239];

   R.~Banerjee, B.~Chakraborty, S.~Ghosh, P.~Mukherjee and S.~Samanta,
  Found.\ Phys.\  {\bf 39}, 1297 (2009)
  [arXiv:0909.1000 [hep-th]];


\bibitem{Banerjee:2009xx}
  R.~Banerjee, S.~Gangopadhyay and S.~K.~Modak,
  Phys.\ Lett.\  B {\bf 686}, 181 (2010)
  [arXiv:0911.2123 [hep-th]].


\bibitem{GUPM}
  H.~Hinrichsen and A.~Kempf,
  J.\ Math.\ Phys.\  {\bf 37}, 2121 (1996)
  [arXiv:hep-th/9510144].

  A.~Kempf,
  J.\ Math.\ Phys.\  {\bf 38}, 1347 (1997)
  [arXiv:hep-th/9602085].
  
  A.~Kempf,
  Phys.\ Rev.\  D {\bf 54}, 5174 (1996)
  [Erratum-ibid.\  D {\bf 55}, 1114 (1997)]
  [arXiv:hep-th/9602119].


\bibitem{Mirza:2009mh}
  B.~Mirza and M.~Zarei,
  Phys.\ Rev.\  D {\bf 79}, 125007 (2009)
  [arXiv:0901.1930 [hep-th]].
  O. Nairz, M. Arndt and A. Zeilinger
  Phys.Rev. A {\bf65}, 032109 (2002).


\bibitem{PhenomenologicalConsequencesNCG}

  J.~L.~Hewett, F.~J.~Petriello and T.~G.~Rizzo,
  Phys.\ Rev.\  D {\bf 64} (2001) 075012
  [arXiv:hep-ph/0010354].

  H.~Arfaei and M.~H.~Yavartanoo,
  arXiv:hep-th/0010244.

  P.~Mathews,
  Phys.\ Rev.\  D {\bf 63} (2001) 075007
  [arXiv:hep-ph/0011332].

  A.~Anisimov, T.~Banks, M.~Dine and M.~Graesser,
  Phys.\ Rev.\  D {\bf 65} (2002) 085032
  [arXiv:hep-ph/0106356].

  S.~Godfrey and M.~A.~Doncheski,
  Phys.\ Rev.\  D {\bf 65} (2002) 015005
  [arXiv:hep-ph/0108268].

  S.~Baek, D.~K.~Ghosh, X.~G.~He and W.~Y.~P.~Hwang,
  Phys.\ Rev.\  D {\bf 64} (2001) 056001
  [arXiv:hep-ph/0103068].

  T.~J.~Konopka and S.~A.~Major,
  New J.\ Phys.\  {\bf 4} (2002) 57
  [arXiv:hep-ph/0201184].

  J.~L.~Hewett, F.~J.~Petriello and T.~G.~Rizzo,
  in {\it Proc. of the APS/DPF/DPB Summer Study on the Future of Particle Physics (Snowmass 2001) } ed. N.~Graf,
  {\it In the Proceedings of APS / DPF / DPB Summer Study on the Future of Particle Physics (Snowmass 2001), Snowmass,
  Colorado, 30 Jun - 21 Jul 2001, pp E3064}[arXiv:hep-ph/0201275].

  Y.~Liao and C.~Dehne,
  Eur.\ Phys.\ J.\  C {\bf 29} (2003) 125
  [arXiv:hep-ph/0211425].

  T.~G.~Rizzo,
  Int.\ J.\ Mod.\ Phys.\  A {\bf 18} (2003) 2797
  [arXiv:hep-ph/0203240].

  I.~Hinchliffe, N.~Kersting and Y.~L.~Ma,
  Int.\ J.\ Mod.\ Phys.\  A {\bf 19} (2004) 179
  [arXiv:hep-ph/0205040].


\bibitem{ChangingInteractionStructureNCG}
  J.~Wess,
  {\it Prepared for 10th International Conference on Supersymmetry and Unification of Fundamental Interactions (SUSY02),    
  Hamburg, Germany, 17-23
  Jun 2002}

  X.~Calmet and M.~Wohlgenannt,
  Phys.\ Rev.\  D {\bf 68} (2003) 025016
  [arXiv:hep-ph/0305027].

  X.~Calmet,
  arXiv:hep-th/0401212.

  X.~Calmet and A.~Kobakhidze,
  Phys.\ Rev.\  D {\bf 72} (2005) 045010
  [arXiv:hep-th/0506157].

  X.~Calmet,
  Eur.\ Phys.\ J.\  C {\bf 50} (2007) 113
  [arXiv:hep-th/0604030].

  X.~Calmet, B.~Jurco, P.~Schupp, J.~Wess and M.~Wohlgenannt,
  Eur.\ Phys.\ J.\  C {\bf 23} (2002) 363
  [arXiv:hep-ph/0111115].

\bibitem{Minwalla:1999px}
  S.~Minwalla, M.~Van Raamsdonk and N.~Seiberg,
  JHEP {\bf 0002} (2000) 020
  [arXiv:hep-th/9912072].


\end{thebibliography}
\end{document}